\documentclass[a4paper]{jpconf}
\usepackage{graphicx}
\begin{document}
\title{Proof of shock-excited H$_2$ in low-ionization structures of PNe}

\author{Stavros Akras$^1$, Denise R. Gon\c{c}alves$^1$ and Gerardo Ramos-Larios$^2$}

\address{$^1$ Observat\'orio do Valongo, Universidade Federal do Rio de Janeiro, Ladeira Pedro Antonio 43, 20080-090, Rio de Janeiro, Brazil\\
$^2$ Instituto de Astronom\'ia y Meteorolog\'ia, Av. Vallarta No. 2602. Col. Arcos Vallarta, CP 44130, Guadalajara, Jalisco, Mexico}

\ead{$^1$ akras@astro.ufrj.br}

\begin{abstract}
We report the detection of near-IR H$_2$ lines emission from low-ionization structures (LISs) in planetary nebulae. The deepest, high-angular resolution H$_2$ 1-0 S(1) at 2.122 $\mu$m, and H$_2$ 2-1 S(1) at 2.248 $\mu$m images of K~4-47 and NGC~7662, obtained using NIRI@Gemini-North, 
are presented here. K~4-47 reveals a remarkable high-collimated bipolar structure. The H$_2$ emission emanates from the walls of the bipolar outflows 
and a pair of knots at the tips of these outflows. The H$_2$ 1-0 S(1)/2-1 S(1) line ratio is $\sim$7-10 which indicates shock interaction due to both the lateral expansion of the gas in the outflows and the high-velocity knots. The strongest line, H$_2$ v=1-0 S(1), is also detected in several LISs located at the periphery of the outer shell of the elliptical PN NGC~7662, whereas only four knots are detected in the H$_2$ v=2-1 S(1) line. These knots have H$_2$ v=1-0 S(1)/v=2-1 S(1) values between 2 and 3. These data confirm the presence of molecular gas in both highly (K~4-47) and slowly moving LISs (NGC~7662). The H$_2$ emission in K~4-47 is powered by shocks, whereas in NGC~7662 is due to photo-ionization by the central star. Moreover, a likely correlation is found between the H$_2$ v=1-0 S(1)/H$_2$ v=2-1 S(1) and [N~{\sc ii}]/H$\alpha$ line ratios.

\end{abstract}

\section{Introduction}

Optical imaging surveys of planetary nebulae (PNe) have revealed that a fraction of PNe possess, besides the large-scale structures such as rims, shells, haloes, some small-scale structures (e.g. \cite{Bal3}, \cite{Cor2}, \cite{Gon01}). These structures 
are prominent in the low-ionization emission lines such as [N~{\sc ii}], [S~{\sc ii}] and [O~{\sc i}] (hereafter LISs). They exhibit a variety of morphologies like knots, jets, and filaments, (\cite{Gon01}), whereas they cover a wide range of expansion velocities
from 30 km s$^{-1}$ up to 350 km s$^{-1}$. Akras \& Gon\c{c}alves (\cite{Akr}), studying a sample of Galactic PNe with LISs, demonstrate that the 
excitation mechanism of these structures is a combination of UV-photons and shocks. The contribution of each mechanism depends on parameters like
the distance of LISs to the central star, the stellar parameters ($\rm{T_{eff}}$, L$\odot$) and LIS's expansion velocity.

The physical properties like $\rm{T_e}$,  $\rm{N_e}$ and chemical abundances of LISs have been studied by different groups (e.g. \cite{Akr}, \cite{Bal1}, \cite{Gon09}), and some of the most important conclusions are: i) there is no 
difference in $\rm{T_e}$ between the LISs and the nebular components, ii) LISs have systematically lower $\rm{N_e}$ compared 
to the surrounding medium (e.g.\cite{Akr}, \cite{Bal1}) and iii) there is no difference in chemical abundances that could provide an explanation for the enhancement of the low-ionization emission lines.  

The formations models of LISs predict that they are denser structures than the surrounding ionized medium. This result is found to be inconsistent 
with the observations. A possible explanation for this discrepancy may be that the formation models refer to the total density of gas 
(dust, atomic and molecular) and not only to the  $\rm{N_e}$. Gon\c{c}alves et al. (\cite{Gon09}) have proposed that LISs may also 
be made of molecular gas. H$_2$ emission has been detected in the cometary knots in the Helix (\cite{Hug}, \cite{Mant}) and knots/filaments in 
NGC 2346 (\cite{Man}). 

Generally, H$_2$ emission has been detected in several PNe (e.g. \cite{Mar}). A comparison of H$_2$ and optical line (e.g. [N~{\sc ii}] and [O~{\sc i}]) images has revealed similar morphologies, suggesting that both emissions emanate from the same regions. A recent theoretical work by Aleman \& Gruenwald (\cite{Ale}) has shown that the peak intensities of the optical low-ionization and H$_2$ lines occur in a narrow transition zone between the ionized and neutral (photo-dissociation) regions.  Moreover, an empirical relation between the fluxes of the [O I] $\lambda$6300 and H$_2$ v=1-0 S(1) lines for Galactic PNe was reported by Reay et al. (\cite{Rea}). Hence, the detection of strong low-ionization lines in LISs may also suggest the presence of H$_2$ gas.


\section{Observations}

The deepest, high-angular resolution H$_2$ v=1-0 S(1) at 2.122 $\mu$m and H$_2$ v=2-1 S(1) at 2.242 $\mu$m images were obtained 
for K~4-47 and NGC~7662 using the NIRI instrument on the Gemini-North 8~m telescope on Manua Kea in Hawaii. 
For these observations, the f/6 configuration (pixel scale=0.117~arcsec and field of view of 120 arcsec) was used. The exposure
times were estimated using the empirical relation from Reay et al. (\cite{Rea}). The final continuum-subtracted images are 
presented in Figures 1 and 2. 

\begin{figure}[h]
\includegraphics[width=24pc]{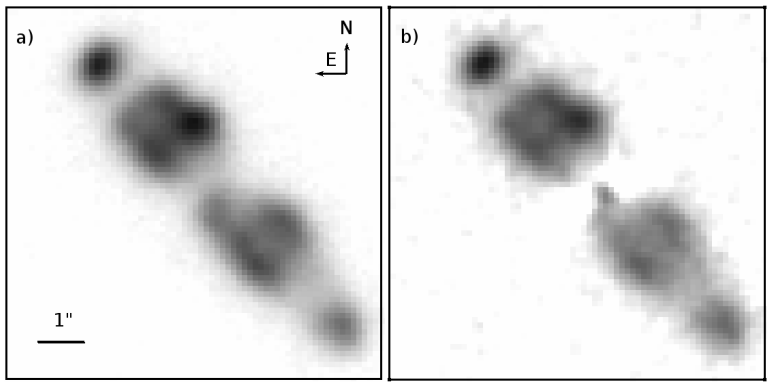}\hspace{2pc}%
\begin{minipage}[b]{11pc}\caption{\label{label}Line images of the bipolar planetary nebula K 4-47. (a) Continuum subtracted H$_2$ v=1-0 S(1), 
(b) H$_2$ v=2-1 S(1) emission lines.} 
\end{minipage}
\end{figure}

\section{Discussion} 

\subsection{The highly collimated bipolar nebula K 4-47}

The narrow-band near-IR H$_2$ images of K 4-47 (Fig.~1) reveal a remarkable high-collimated bipolar structure with the emission arising 
from the walls of the bipolar outflows. At the tips of these outflows lies a pair of fast-moving, low-ionization knots (100~km~s$^{-1}$; \cite{Cor}). 

These knots exhibit very strong [N~{\sc i}]~$\lambda$5200 and [O~{\sc i}]~$\lambda$6300 emission lines that are usually attributed 
to high-velocity shocks. Gon\c{c}alves et al. (\cite{Gon04}), running a number of shock models, came to the conclusion that the 
expansion velocity of these knots must be up to 250--300~km~s$^{-1}$ in order to reproduce these lines. Despite their very high velocities, 
their H$_2$ v=1-0 S(1)/H$_2$ v=2-1 S(1) line ratio is found to be $\sim$7, lower than the typical value of 10 for shock-excited regions (\cite{bvd}). 

Regarding the bipolar outflows, a comparison between our near-IR and optical images from \cite{Cor} illustrates that the ionized gas (optical emission) is concentrated in a inner highly collimated structure, surrounded by the H$_2$ outflows. This structure of K~4-47 completely resembles that 
of the M~2-9 and CRL~618 PNe indicating a possible connection among these objects. The large H$_2$ v=1-0 S(1)/H$_2$ v=2-1 S(1) line ratio, 
estimated in the bipolar shell, was unexpected. Although, the recent hydrodynamic models by Balick B. and collaborators (\cite{Bal2}) can 
adequately explain our findings. The H$_2$ emission from the bipolar outflows can be explained as the result of the interactions between a jet or 
bullet with the AGB material. As the jet/bullet moves through the AGB material forms an conical structure that expands laterally outward with a 
velocity that increases with the distance from the central star. At the same time, the jet/bullet continues moving outwards, with a velocity 
proportional to the distance from the central star, dissociating the AGB H$_2$ gas, 
which later is being ionized by UV-photons (optical images). The surface brightnesses in the H$_2$ v=1-0 S(1) and H$_2$ v=2-1 S(1) emission 
lines are found to range from 0.2 to 1$\times$10$^{-15}$~erg~cm$^{-1}$~s$^{-1}$~arc$^{-2}$ and from 0.2 to 1 $\times$10$^{-16}$~erg~cm$^{-1}$~s$^{-1}$~arc$^{-2}$, respectively.

\begin{figure}[h]
\includegraphics[width=35pc]{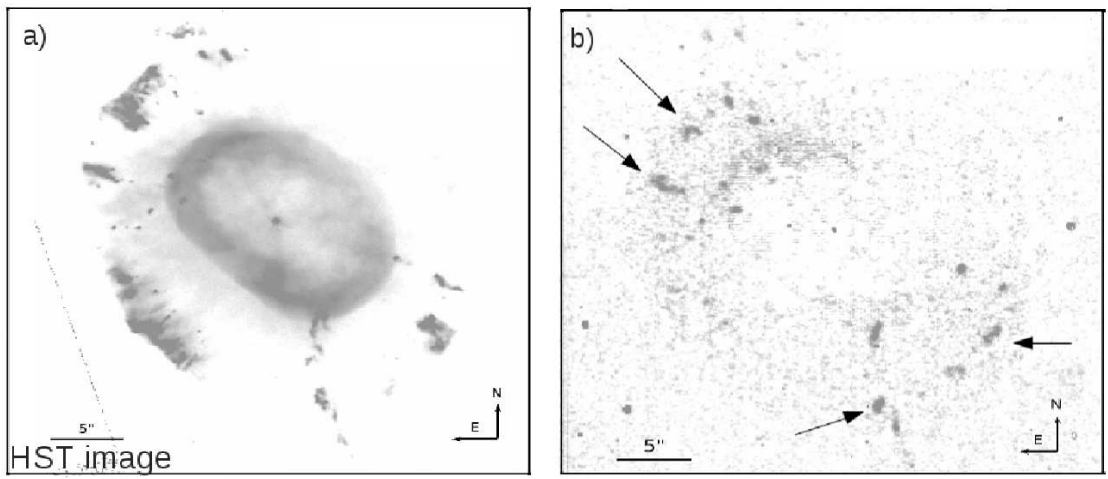}\hspace{2pc}
\caption{\label{label}Continuum subtracted H$_2$ v=1-0 S(1) line image of the elliptical planetary nebula 
NGC~7662. The arrows indicate the four LISs in which the H$_2$ v=2-1 S(1) emission is detected.} 
\end{figure}

\subsection{The elliptical nebula NGC 7662}

NGC 7662 is an elliptical PN that posses almost two dozens of LISs (knots and a jet-like structure) embedded in the outer shell with expansion 
velocities that vary from 30 to 70 km s$^{-1}$ (see \cite{Per}). All these structures are easily discerned in the [N~{\sc ii}] line image (\cite{Gue}), whereas spectroscopic data have revealed that they also exhibit a strong [O~{\sc i}]~$\lambda$6300 line (\cite{Per},\cite{Gon09}). Despite that the former line is a strong indicator of shock interactions, most of the LISs have low velocities except from the southern jet-like feature. 
(see \cite{Per}). 

Here, we present the deepest near-IR images of this nebula. The H$_2$ v=1-0 S(1) emission line is detected in almost all optically identified LISs 
(Fig.~2, right panel.), whereas only four LISs are found to have H$_2$ v=2-1 S(1) emission. The former line is 
found to range from 1 to 4.8 $\times$ 10$^{-16}$~erg~cm$^{-1}$~s$^{-1}$~arc$^{-2}$, whereas the latter from 0.6 to 1 $\times$ 10$^{-16}$~erg~cm$^{-1}$~s$^{-1}$~arc$^{-2}$. No H$_2$ emission is found associated with the nebular shells. This confirms that LISs 
are likely molecular condensations embedded in a nebula. The value of H$_2$ v=1-0 S(1)/H$_2$ v=2-1 S(1) line ratio is found to range between 3 and 5. These values imply that the LISs in this nebula are predominantly photo-ionized. Nevertheless, shocks cannot be discarded.

\begin{figure}[h]
\includegraphics[width=21pc]{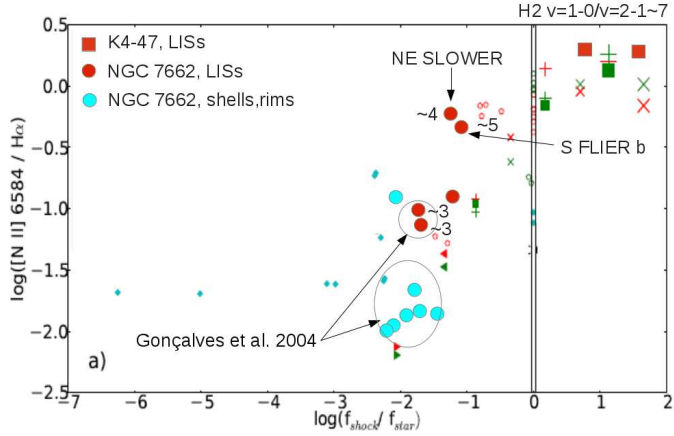}\hspace{2pc}%
\begin{minipage}[b]{15pc}\caption{\label{label} Plot of [N II]/H$\alpha$ vs. log($\rm{f_{shocks}}$ / $\rm{f_{star}}$) for K 4-47 (\cite{Gon09}) and 
NGC 7662 (\cite{Per}) from the diagnostic diagram of Akras \& Gon\c{c}alves (\cite{Akr}) for photo-ionized and shock-excited regions. The numbers 
indicate the value of the H$_2$ v=1-0 S(1)/H$_2$ v=2-1 S(1) line ratio from six LISs (the pair of knots in K 4-47 and the four knots in NGC~7662).}
\end{minipage}
\end{figure}

\section{Conclusion}

New, deep, high-angular resolution near-IR H$_2$ images confirmed the presence of molecular gas in fast- (K 4-47) and slow-moving LISs 
(NGC~7662). H$_2$ emission was also detected in the dense walls of the bipolar outflows of K 4-47 that implies a lateral expansion 
in agreement with the predictions from hydrodynamic models. Important morphological similarities among K~4-47, M~2-9 and CRL~618 have been found. 

In the NGC~7662 nebula, H$_2$ emission was detected in several LISs, while it is totally absent in the nebular shells. 
Using the new diagnostic diagram from Akras \& Gon\c{c}alves (\cite{Akr}), and the spectroscopic data of the six LISs with measured H$_2$ v=1-0 S(1)/H$_2$ v=2-1 S(1) line ratio, we found that the near-IR line ratio increases with the [N~{\sc ii}]/H$\alpha$ optical line ratio (Fig.~3). This result may reflect a similar origin for the near-IR and optical line ratios. In conclusion, the H$_2$ emission in K~4-47 is powered by shocks, whereas in NGC~7662 is due 
to photo-ionization by the central star with a possible contribution of shocks.

\ack
Based on observations obtained at the Gemini Observatory, which is operated by the Association of Universities or Research in Astronomy, Inc., under a cooperative agreement with the NSF on behalf of the Gemini partnership. This work is supported by CAPES (program A35/2013) and FAPERJ (program E-26/111.817/2012).

\end{document}